\documentclass[11pt]{article}
\usepackage{amsmath,amsfonts,amssymb}
\numberwithin{equation}{section}

\newcommand{\nc}{\newcommand}
\nc{\bib}{\bibitem}
\nc{\al}{\alpha}
\nc{\g}{\gamma}
\nc{\G}{\Gamma}
\nc{\D}{\Delta}
\nc{\eps}{\epsilon}
\nc{\vareps}{\varepsilon}
\nc{\la}{\lambda}
\nc{\La}{\Lambda}
\nc{\var}{\varphi}
\nc{\pa}{\partial}
\nc{\nn}{\nonumber \\ }
\nc{\be}{\begin{equation}}
\nc{\ee}{\end{equation}}
\nc{\bea}{\begin{eqnarray}}
\nc{\eea}{\end{eqnarray}}
\nc{\bra}[1]{\langle {#1}|}
\nc{\ket}[1]{|{#1}\rangle}
\nc{\sbar}{\bar{s}}
\nc{\Ab}{\bar{A}}
\nc{\Db}{\bar{D}}
\nc{\Lc}{\mathcal{L}}
\nc{\Oc}{\mathcal{O}}
\nc{\Qh}{\hat{Q}}
\nc{\ab}{\bar{a}}
\nc{\kappab}{\bar{\kappa}}
\nc{\Xib}{\bar{\Xi}}
\nc{\kb}{\bar{k}}
\nc{\rb}{\bar{r}}
\nc{\Rb}{\bar{R}}
\nc{\kh}{\hat{k}}
\nc{\rh}{\hat{r}}
\nc{\sh}{\hat{s}}
\nc{\thh}{\hat{t}}
\nc{\phih}{\hat{\phi}}
\nc{\rhoh}{\hat{\rho}}
\nc{\Dh}{\hat{\Delta}}
\nc{\Kc}{\mathcal{K}}

\begin{document}

\topmargin -5mm
\oddsidemargin 5mm

\setcounter{page}{1}

\vspace{8mm}
\begin{center}
{\LARGE {\bf On the CFT duals for near-extremal black holes}}

\vspace{8mm}
 {\LARGE J{\o}rgen Rasmussen}
\\[.3cm]
 {\em Department of Mathematics and Statistics, University of Melbourne}\\
 {\em Parkville, Victoria 3010, Australia}
\\[.4cm]
 {\tt j.rasmussen@ms.unimelb.edu.au}

\end{center}

\vspace{8mm}
\centerline{{\bf{Abstract}}}
\vskip.4cm
\noindent
We consider Kerr-Newman-AdS-dS black holes near extremality and work out
the near-horizon geometry of these near-extremal black holes. 
We identify the exact $U(1)_L\times U(1)_R$ isometries 
of the near-horizon geometry and
provide boundary conditions enhancing them to a pair of commuting Virasoro algebras.
The conserved charges of the corresponding asymptotic symmetries are found to be
well defined and non-vanishing and to yield central charges $c_L\neq0$ 
and $c_R=0$. The Cardy formula subsequently reproduces the
Bekenstein-Hawking entropy of the black hole. This suggests that the near-extremal
Kerr-Newman-AdS-dS black hole is holographically dual to a non-chiral two-dimensional
conformal field theory.
\renewcommand{\thefootnote}{\arabic{footnote}}
\setcounter{footnote}{0}

\newpage

\section{Introduction}

In the seminal work \cite{BH86}, quantum gravity on three-dimensional anti-de Sitter (AdS) space
was found to be holographically dual to a two-dimensional conformal field theory (CFT).
The Kerr/CFT correspondence \cite{GHSS0809} extends this duality by
asserting that quantum theory on the near-horizon geometry of the extremal four-dimensional
Kerr black hole \cite{BH9905} is holographically dual to a chiral CFT in two dimensions.
This near-horizon geometry has a $U(1)_L\times SL(2,\mathbb{R})_R$ isometry group 
and is called the NHEK geometry, and using the formalism of~\cite{BB0111}, 
it was found \cite{GHSS0809} that certain boundary conditions enhance
the $U(1)_L$ symmetry to a Virasoro algebra. The consistency of these boundary conditions was
subsequently confirmed in \cite{AHMR0906,DRS0906}, while 
a partial classification of boundary conditions with the same properties was provided 
in \cite{Ras0908}. Boundary conditions enhancing the 
$SL(2,\mathbb{R})_R$ isometries to a Virasoro algebra were examined in
\cite{MTY0907,Ras0908}.
Strong evidence \cite{GHSS0809} for the Kerr/CFT correspondence is found in the exact
agreement between the macroscopic Bekenstein-Hawking entropy \cite{Bek73}
of the black hole and the Cardy formula for the dual CFT entropy.
This analysis has been successfully generalized and applied to a variety of extremal
black holes \cite{wake1,CCLP0812,wake2,Wen0903,AHMR0906,DRS0906,wake3}.
The extension to the near-horizon geometry of the extremal Kerr-Newman-AdS-dS black hole
\cite{Car68,HMNS0811}, in particular, is worked out in \cite{HMNS0811}.

Extending the Kerr/CFT correspondence to the {\em near}-extremal Kerr black hole, however,
has presented some serious challenges. The dual two-dimensional CFT should be non-chiral
and hence accommodate two mutually commutative Virasoro algebras. 
Based on a variety of approaches, several insightful results on these near-extremal black 
holes have been obtained \cite{CaLa0908}.
Extending the approach of \cite{MS9609}, in particular,
the work \cite{BHSS0907} on black-hole superradiance and the subsequent 
generalizations thereof \cite{CL0908} have provided highly non-trivial
support for the Kerr/CFT correspondence.
These results, however, are all based on deviations from the
geometrical approach of \cite{GHSS0809} since consistent boundary conditions
which allow for both left- and right-moving sectors have been identified only very recently
\cite{Ras1004}. Another very recent development is the analysis of the so-called hidden conformal 
symmetry \cite{CMS1004} of the Kerr black hole.
It is noted that this symmetry is not a symmetry of the spacetime geometry.

It is the main objective of the present work to extend the work \cite{Ras1004} on the 
near-horizon geometry of the near-extremal Kerr black hole, the so-called near-NHEK geometry, 
to the near-horizon geometry of the near-extremal Kerr-Newman-AdS-dS black hole using 
the results \cite{HMNS0811} on the extremal case.
We work out the metric of this near-horizon geometry, identify its exact $U(1)_L\times U(1)_R$ 
isometry group, and provide boundary conditions enhancing these symmetries to a pair of 
commuting Virasoro algebras.
The conserved charges of the corresponding asymptotic symmetries are found to be
well defined and non-vanishing and to yield central charges $c_L\neq0$ 
and $c_R=0$. The Cardy formula subsequently reproduces the
Bekenstein-Hawking entropy of the black hole. This supports the assertion that the near-extremal
Kerr-Newman-AdS-dS black hole is holographically dual to a {\em non-chiral} two-dimensional
CFT. Here we only consider the geometry of the Kerr-Newman-AdS-dS black hole but
hope to address elsewhere the inclusion of the gauge field.

It is noted that the Virasoro algebra has been observed in other black-hole contexts as well.
Extending the work~\cite{Str9712} on the entropy of three-dimensional black holes such as the
BTZ black hole~\cite{BTZ9204}, it was found in~\cite{Car9812,Sol9812} 
that a copy of the Virasoro algebra appears in the near-horizon region of any black hole
and that it reproduces the black-hole entropy using the Cardy formula.

\section{Kerr-Newman-AdS-dS black hole}

\subsection{Geometry}

In Boyer-Lindquist-type coordinates, and using the unit convention where $G=\hbar=c=1$, 
a Kerr-Newman-AdS-dS black hole \cite{Car68,HMNS0811} is described by
\be 
 d\sh^2=-\frac{\D_r}{\rho^2}\Big(d\thh-\frac{a}{\Xi}\sin^2\theta d\phih\Big)^2
  +\frac{\rho^2}{\D_r}d\rh^2+\frac{\rho^2}{\D_\theta}d\theta^2
  +\frac{\D_\theta}{\rho^2}\sin^2\theta\Big(ad\thh-\frac{\rh^2+a^2}{\Xi}d\phih\Big)^2
\label{ds2}
\ee
where
\bea
 &&\D_r=(\rh^2+a^2)\big(1+\frac{\rh^2}{\ell^2}\big)-2M\rh+q^2,\qquad
  q^2=q_e^2+q_m^2\nn
 &&\D_\theta=1-\frac{a^2}{\ell^2}\cos^2\theta,\qquad
  \rho^2=\rh^2+a^2\cos^2\theta,\qquad
  \Xi=1-\frac{a^2}{\ell^2}
\eea
This metric is parameterized by the ADM mass $M_{ADM}=M/\Xi^2$, 
angular momentum $J=aM/\Xi^2$
and electric and magnetic charges $Q_e=q_e/\Xi$ and $Q_m=q_m/\Xi$, and
it is AdS (dS) for positive (negative) renormalized cosmological constant $\ell^{-2}$. 
It reduces to Kerr-Newman for
$\ell^{-2}=0$ and further to Kerr for $q^2=0$.
The horizons are located at the positive zeros of $\D_r$, and the value of $\rh$ at the outer
horizon is denoted by $r_+$.
The Hawking temperature $T_H$, angular velocity $\Omega_H$ of the event horizon,
and entropy $S$ are given by
\be
 T_H=\frac{r_+\big(1+\frac{a^2}{\ell^2}+\frac{3r_+^2}{\ell^2}-\frac{a^2+q^2}{r_+^2}
    \big)}{4\pi(r_+^2+a^2)},\qquad
 \Omega_H=\frac{a\Xi}{r_+^2+a^2},\qquad
 S=\frac{\pi(r_+^2+a^2)}{\Xi}
\label{TH}
\ee
For later convenience, we follow \cite{HMNS0811} and introduce
\be
 r_0^2=\frac{(r_+^2+a^2)\big(1-\frac{r_+^2}{\ell^2}\big)}{1+\frac{6r_+^2}{\ell^2}
  -\frac{3r_+^4}{\ell^4}-\frac{q^2}{\ell^2}},\qquad
  \rho_+^2=r_+^2+a^2\cos^2\theta
\label{rrho}
\ee
and
\be
 \G(\theta)=\frac{\rho_+^2 r_0^2}{r_+^2+a^2},\qquad
 \al(\theta)=\frac{r_+^2+a^2}{\D_\theta r_0^2},\qquad
 \g(\theta)=\frac{\D_\theta\sin^2\theta(r_+^2+a^2)^2}{\Xi^2\rho_+^2},\qquad
 k=\frac{2a\Xi r_0^2r_+}{(r_+^2+a^2)^2}
\label{Gagk}
\ee

\subsection{Extremal black holes}

When the inner and outer horizons coalesce, the black hole is said to be extremal.
This happens when the otherwise single pole $r_+$ of $\D_r$ is a double pole in which case
\be
 \D_r(r_+)=\D'_r(r_+)=0
\label{DhDh}
\ee
We denote the value of $\rh$ at this single horizon by $\rb$ and use the related notation
$\ab$, $\bar{T}_H$ and $\bar{\Omega}_H$, in particular, for the evaluation of the various
parameters at extremality.
The conditions (\ref{DhDh}) can be used to express $M$ and $\ab^2$ in terms
of $\rb$ 
\be
 M=\frac{\rb\big[\big(1+\frac{\rb^2}{\ell^2}\big)^2-\frac{q^2}{\ell^2}\big]}{1-\frac{\rb^2}{\ell^2}},
   \qquad
 \ab^2=\frac{\rb^2\big(1+\frac{3\rb^2}{\ell^2}\big)-q^2}{1-\frac{\rb^2}{\ell^2}}
\ee
and we readily recover $\bar{T}_H=0$.

\subsection{Near-horizon geometry of extremal black holes}

To describe the near-horizon geometry of the extremal Kerr-Newman-AdS-dS black hole,
we follow \cite{BH9905,HMNS0811} and introduce the new coordinates $t,r,\phi$ where
\be
 \thh=\frac{t\rb_0}{\eps},\qquad 
 \rh=\rb+\eps\rb_0 r,\qquad 
 \phih=\phi+\frac{\bar{\Omega}_H\rb_0t}{\eps}
\ee
The near-horizon geometry is obtained by taking the limit $\eps\to0$ in which case
the metric (\ref{ds2}) becomes
\be
 d\bar{s}^2=\bar{\Gamma}(\theta)\Big(-r^2dt^2+\frac{dr^2}{r^2}+\bar{\al}(\theta)d\theta^2\Big)
  +\bar{\gamma}(\theta)\big(d\phi+\kb rdt\big)^2
\label{ds2Gamma}
\ee
The exact isometry group of the near-horizon geometry (\ref{ds2Gamma}) 
is $U(1)_L\times SL(2,\mathbb{R})_R$ and it is generated by
\be
 \big\{\pa_\phi\big\}\cup\big\{\pa_t,\ t\pa_t-r\pa_r,\ \big(t^2+\frac{1}{r^2}\big)\pa_t
   -2tr\pa_r-\frac{2\kb}{r}\pa_\phi\big\}
\label{iso}
\ee

\subsection{Near-extremal geometry}

We are interested in infinitesimal excitations above extremality of the
Kerr-Newman-AdS-dS black hole. To describe this near-extremal geometry, we consider the
situation where the outer horizon $r_+$ is infinitesimally larger than the extremal
horizon $\rb$
\be
 r_+=\rb+\eps\kappab   
\ee
Here $\kappab$ is expressed in terms of the parameters of the extremal black hole, 
while $\eps$ is an infinitesimal parameter whose limit $\eps\to0$ we are ultimately interested in,
where
\be
 a^2=\ab^2-\frac{\eps^2\kappab^2(\rb^2+\ab^2)}{\rb_0^2\big(1+\frac{\rb^2}{\ell^2}\big)}
    +\Oc(\eps^3),\qquad
 T_H=\frac{\eps\kappab}{2\pi\rb_0^2}+\Oc(\eps^2),\qquad
 \Omega_H=\bar{\Omega}_H\big(1-\frac{2\eps\kappab\rb}{\rb^2+\ab^2}\big)+\Oc(\eps^2)
\ee
Following \cite{CCLP0812}, we introduce
\be
 T_L=
 \lim_{r_+\to\rb}\frac{T_H}{\bar{\Omega}_H-\Omega_H}
  =\frac{1}{2\pi\kb}
\ee
and interpret it as the Frolov-Thorne temperature \cite{FT89} associated with the azimuthal
angle $\phi$. Although the asymptotically-defined Hawking temperature $T_H$ vanishes in
the limit $\eps\to0$, we can use that $T_H/\eps$ is non-vanishing and finite to introduce
\be
 T_R=\lim_{r_+\to\rb}\frac{r_0^2 T_H}{r_+\eps}=\frac{\kappab}{2\pi \rb}
\ee
and interpret it as a temperature associated with the near-extremality of the geometry.

\subsection{Near-horizon geometry of near-extremal black holes}

To describe the near-horizon geometry of the near-extremal Kerr-Newman-AdS-dS black hole,
we use the coordinates $t,r,\phi$ where
\be
 \thh=\frac{\rb_0^2t}{\eps},\qquad 
 \rh=\rb+\eps(r+\la\kappab),\qquad 
 \phih=\phi+\frac{\bar{\Omega}_H\rb_0^2t}{\eps}
\label{trphi}
\ee
and where we have introduced the additional parameter $\la$.
In the limit $\eps\to0$, we find
\be
 \D_r=\frac{\eps^2(\rb^2+\ab^2)}{\rb_0^2}\big(r+(\la-1)\kappab\big)
   \big(r+(\la+1)\kappab\big)+\Oc(\eps^3)
\ee
and hence
\bea
 ds^2&=&\Gamma(\theta)\Big[-\big(r+(\la-1)\kappab\big)
   \big(r+(\la+1)\kappab\big)dt^2
  +\frac{dr^2}{\big(r+(\la-1)\kappab\big)
   \big(r+(\la+1)\kappab\big)}
  +\al(\theta)d\theta^2\Big]\nn
 &+&\g(\theta)\big[d\phi+k(r+\la\kappab)dt\big]^2
\label{ds2lambda}
\eea
For all $\la$, this is invariant under the action of the exact isometry group $U(1)_L\times U(1)_R$
generated by 
\be
 \big\{\pa_\phi\big\}\cup\big\{\pa_t\big\}
\ee 
These isometries commute with translations in $r$ generated by $\pa_r$ and realized by
varying $\la$. We note that the determinant of the metric (\ref{ds2lambda}) only depends on
$\theta$ as
\be
 \sqrt{-g}=\sqrt{\G^3(\theta)\al(\theta)\g(\theta)}=\frac{\sin\theta\rho_+^2 r_0^2}{\Xi}
\ee
In the following, focus will be on $\la=1$ in which case
\be
 r=\frac{\rh-r_+}{\eps}
\ee
and
\be
 ds^2\big|_{\la=1}=\Gamma(\theta)\Big(-r(r+2\kappab)dt^2
  +\frac{dr^2}{r(r+2\kappab)}
  +\al(\theta)d\theta^2\Big)
 +\g(\theta)\big(d\phi+k(r+\kappab)dt\big)^2
\label{ds21}
\ee

As above, special cases are obtained by setting $\ell^{-2}=0$ or $q^2=0$.
The near-NHEK geometry \cite{Wen0903,AHMR0906,BHSS0907,Ras1004}, in particular,
follows when both of these conditions apply. We note that the coordinate convention
for the near-NHEK geometry used in \cite{BHSS0907,Ras1004} differs from the one following 
from (\ref{trphi}), but is obtained by the replacement $t,r\to t/M,Mr$.

\section{Holographically dual CFT}

\subsection{Boundary conditions and conserved charges}

We are interested in perturbations $h_{\mu\nu}$
of the near-horizon geometry (\ref{ds21}) of the near-extremal Kerr-Newman-AdS-dS
black hole.
Following \cite{Ras1004} on the near-NHEK geometry, we introduce the fall-off conditions
\be
 h_{\mu\nu}=\Oc\!\left(\!\!\begin{array}{cccc} 
   r^2&r^{-3}&r^{1}&r^{-2} \\ 
   &r^{-4}&r^{-1}&r^{-3} \\ 
   &&1&r^{-2} \\ 
   &&&r^{-2} \end{array} \!\!\right),
  \qquad\quad h_{\mu\nu}=h_{\nu\mu}
\label{hnear}
\ee
in addition to a couple of supplementary conditions to be discussed below.
Asymptotic symmetries are generated by the diffeomorphisms whose action on the
metric generates metric fluctuations compatible with the full set of boundary conditions.
Initially focusing on the fall-off conditions (\ref{hnear}), we find the asymptotic Killing vectors
\bea
 K_\eps&=&\big[\Oc(r^{-4})\big]\pa_t+\big[-(r+\kappab)\eps'(\phi)+\Oc(r^{-1})\big]\pa_r
   +\big[\eps(\phi)+\Oc(r^{-3})\big]\pa_\phi
   +\big[\Oc(r^{-2})\big]\pa_\theta   \nn
 \Kc_\vareps&=&\big[\vareps(t)+\Oc(r^{-4})\big]\pa_t+\big[\Oc(r^{-1})\big]\pa_r
   +\big[\Oc(r^{-3})\big]\pa_\phi
   +\big[\Oc(r^{-2})\big]\pa_\theta
\eea
where $\eps(\phi)$ and $\vareps(t)$ are smooth functions.
The generators of the corresponding asymptotic symmetries thus read
\be
 \xi=-(r+\kappab)\eps'(\phi)\pa_r+\eps(\phi)\pa_\phi,\qquad
 \zeta=\vareps(t)\pa_t
\label{xizeta}
\ee
and form a commuting pair of centreless Virasoro algebras
\be
 \big[\xi_\eps,\xi_{\hat{\eps}}\big]=\xi_{\eps\hat{\eps}'-\eps'\hat{\eps}},\qquad
 \big[\zeta_\vareps,\zeta_{\hat{\vareps}}\big]=\zeta_{\vareps\hat{\vareps}'-\vareps'\hat{\vareps}},
     \qquad
 \big[\xi_\eps,\zeta_\vareps\big]=0
\ee
Along $\xi$ and $\zeta$, the Lie derivatives of the near-horizon geometry (\ref{ds21}) 
are worked out to be
\bea
 \mathcal{L}_\xi g_{\mu\nu}&=&\left(\begin{array}{cccc}
  2(\G-k^2\g)(r+\kappab)^2\eps'(\phi)&0&0&0\\[.2cm]
  0&\frac{2\kappab^2\G\eps'(\phi)}{r^2(r+2\kappab)^2}
    &-\frac{\G(r+\kappab)\eps''(\phi)}{r(r+2\kappab)}&0\\[.2cm]
  0&-\frac{\G(r+\kappab)\eps''(\phi)}{r(r+2\kappab)}&2\g\eps'(\phi)&0\\[.2cm]
  0&0&0&0
  \end{array}\right)\nn
 \mathcal{L}_\zeta g_{\mu\nu}&=&\vareps'(t)\left(\begin{array}{cccc}
 -2\G r(r+2\kappab)+2k^2\g(r+\kappab)^2&0&k\g(r+\kappab)&0\\[.2cm]
  0&0&0&0\\[.2cm]
  k\g(r+\kappab)&0&0&0\\[.2cm]
  0&0&0&0
  \end{array}\right)
\label{Lie}
\eea
where we have introduced the abbreviations
\be
 \G=\G(\theta),\qquad \al=\al(\theta),\qquad \g=\g(\theta)
\ee

To an asymptotic symmetry generator $\eta$, one associates~\cite{BB0111}
the conserved charge
\be
 Q_\eta=\frac{1}{8\pi}\int_{\pa\Sigma}\sqrt{-g}k_\eta[h;g]
   =\frac{1}{8\pi}\int_{\pa\Sigma}\frac{\sqrt{-g}}{4}\eps_{\al\beta\mu\nu}d_\eta^{\mu\nu}[h;g]
     dx^\al\wedge dx^\beta
\label{Q}
\ee
where
\be
 d_\eta^{\mu\nu}[h;g]
  =\eta^\nu D^\mu h-\eta^\nu D_\sigma h^{\mu\sigma} +\eta_\sigma D^\nu h^{\mu\sigma}
  -h^{\nu\sigma}D_\sigma\eta^\mu+\frac{1}{2}hD^\nu\eta^\mu
  +\frac{1}{2}h^{\sigma\nu}\big(D^\mu\eta_\sigma+D_\sigma\eta^\mu\big)
\label{dmunu}
\ee
and where $\pa\Sigma$ is the boundary of a three-dimensional
spatial volume near spatial infinity. Indices are lowered and raised
using the background metric $g_{\mu\nu}$ and its inverse, 
$D_\mu$ denotes a background covariant derivative, 
while $h$ is defined as $h=g^{\mu\nu}h_{\mu\nu}$.
To be a well-defined charge in the asymptotic limit, 
the underlying integral must be finite as $r\to\infty$.
If the charge vanishes, the asymptotic symmetry is rendered trivial.
The asymptotic symmetry group is then generated by the diffeomorphisms 
whose charges are well-defined and non-vanishing.
The algebra generated by the set of well-defined charges is governed
by the Dirac brackets computed~\cite{BB0111} as
\be
 \big\{Q_\eta,Q_{\hat\eta}\big\}
  =Q_{[\eta,\hat\eta]}+\frac{1}{8\pi}\int_{\pa\Sigma}\sqrt{-g}k_{\eta}[\Lc_{\hat\eta}g;g]
\label{QQ}
\ee
where the integral yields the eventual central extension.

Now, considering the asymptotic $r$-expansion of $k_{\pa_t}[h;g]$, we find that its
divergent term (linear in $r$) is independent of $\kappab$ and that its constant term 
(independent of $r$) is at most linear in $\kappab$. 
Subleading terms in $r$ can be ignored as $r\to\infty$. 
We thus follow \cite{Ras1004} and impose the condition
\be
 \big(1-\kappab\pa_{\kappab}\big)Q_{\pa_t}=0
\label{aQ0}
\ee
where $\pa_{\kappab}$ is a formal derivative in $\kappab$.
Furthermore, the conserved charge $Q_{\zeta}$ must be well-defined for all $\zeta$, 
not just for $\vareps(t)=1$ as in (\ref{aQ0}). To understand the potential 
complications arising when extending from $Q_{\pa_t}$ to general $Q_\zeta$,
we examine the difference in their integrands and
find the remarkably simple relation
\be
 \sqrt{-g}\big(k_\zeta[h;g]-\vareps(t)k_{\pa_t}[h;g]\big)\big|_{d\phi\wedge d\theta}=
  -\frac{1}{4}\sqrt{\frac{\al\g}{\G}}\vareps'(t)(r+\kappab)h_{r\phi}
\label{kk}
\ee
valid for all $r$.
It follows, in particular, that the divergent part of the asymptotic $r$-expansion
of $k_\zeta[h;g]$
matches the divergent part of $\vareps(t)k_{\pa_t}[h;g]$ where
\bea
 &&\sqrt{-g}k_{\pa_t}[h;g]\big|_{d\phi\wedge d\theta}
 =\frac{1}{4}\sqrt{\frac{\al}{\G^3\g}}\Big\{-k^2\g^2r^{-1}h_{tt}
  +2k\g\big[k^2\g-\G r\pa_r\big]h_{t\phi}\nn
  &&\hspace{3.3cm} +2\G\big[\G-k^2\g\big]r^2\pa_\phi h_{r\phi}
   +\big[2\G^2-k^2\G\g-k^4\g^2\big]rh_{\phi\phi}\Big\}+\Oc(r^0)
\label{div}
\eea 
Here we have used that $\pa_r h_{\phi\phi}=\Oc(r^{-2})$.
Since the right-hand side of (\ref{kk}) may violate the Jacobi identity for the Dirac brackets of the
conserved charges $Q_\zeta$, we require that $h_{r\phi}$ is a total $\phi$-derivative 
on the boundary
\be
 h_{r\phi}\big|_{\pa\Sigma}=\pa_\phi H_{r\phi}
\label{hH}
\ee
where it is noted that total $\phi$-derivatives can be ignored in the computation of the conserved charge 
$Q_\zeta$. Under the constraints (\ref{hnear}), (\ref{aQ0}) and (\ref{hH}), only perturbations $h$ 
preserving them and only background metrics $g$ which can be reached from the near-extremal 
geometry via a path of such perturbations are considered. 
This should ensure that $Q_\xi$ and $Q_\zeta$ are well defined.

Taking the constraint (\ref{aQ0}) explicitly into account, we find that $Q_{\zeta}$ can be written as
\be
 Q_\zeta=\frac{1}{8\pi}\int_{\pa\Sigma}\sqrt{-g}\kh_\zeta[h;g]
\label{Qzeta}
\ee
where
\be
 \sqrt{-g}\kh_\zeta[h;g]
  =\frac{k^2\kappab\vareps(t)}{2}\sqrt{\frac{\al\g^3}{\G^3}}
   \Big(\frac{h_{tt}}{r^2}-\frac{kh_{t\phi}}{r}\Big)
     d\phi\wedge d\theta+\ldots
\label{kh}
\ee
The dots indicate that terms not contributing to the charge (\ref{Qzeta}) have been omitted.
The conserved charge $Q_\zeta$ is thus non-vanishing for general boundary conditions
satisfying the constraints (\ref{hnear}), (\ref{aQ0}) and (\ref{hH}).
We also find that the integrand in (\ref{Q}) for the conserved charge $Q_\xi$ is given by
\bea
 &&\sqrt{-g}k_\xi[h;g]= \frac{1}{4}\sqrt{\frac{\al}{\G^3\g}}\Big(
   \g\eps(\phi)\big[
 -\frac{k\g h_{tt}}{r^2}
 +\frac{2(k^2\g-\G r\pa_r)h_{t\phi}}{r}
 -2k\G r\pa_\phi h_{r\phi}
  -k(\G+k^2\g)h_{\phi\phi}
 \big]\nn
 &&\hspace{4.5cm}
 +2\G\eps'(\phi)\big[
  k\g rh_{r\phi}
  -\frac{\G\pa_\phi h_{t\phi}}{r}\big]
 +\eps''(\phi)\big[\frac{\G^2h_{t\phi}}{r}\big]
 \Big)d\phi\wedge d\theta+\ldots
\eea
It is recalled that the near-horizon geometry of the extremal Kerr-Newman-AdS-dS 
black hole is obtained by setting $\kappab=0$ in which
case the conserved charge $Q_\zeta$ is seen to vanish.
The conserved charge $Q_\xi$, on the other hand, is unaffected by setting $\kappab=0$.

\subsection{Central charges and entropy}

Following the recipe used in \cite{GHSS0809}, for example, we introduce the basis
\be
 \eps_n(\phi)=-e^{-in\phi}
\ee
and work out the corresponding Dirac bracket algebra (\ref{QQ}). We thus find
\be
 i\{Q_{\xi_n},Q_{\xi_m}\}=(n-m)Q_{\xi_{n+m}}+\frac{c_L}{12}n(n^2-1+b)\delta_{n+m,0}
\ee
where $b$ is a constant while the central charge is given by
\be
 c_L=3k\int_0^\pi\sqrt{\G(\theta)\al(\theta)\g(\theta)}d\theta
\label{cL}
\ee
After introducing the quantum generators
\be
 L_n=Q_{\xi_n}+\frac{bc_L}{24}\delta_{n,0}
\ee
and performing the usual substitution $\{.,.\}\to-i[.,.]$ of Dirac brackets by quantum 
commutators (recalling that we are using the unit convention $\hbar=1$),
the quantum charge algebra is recognized as the centrally-extended Virasoro  algebra
\be
 [L_n,L_m]=(n-m)L_{n+m}+\frac{c_L}{12}n(n^2-1)\delta_{n+m,0}
\ee
With the expressions (\ref{rrho}) and (\ref{Gagk}) appearing in the near-horizon geometry 
(\ref{ds21}) of the near-extremal Kerr-Newman-AdS-dS black hole, the central charge is 
worked out to be
\be
 c_L=\frac{12r_+\sqrt{\big(r_+^2+\frac{3r_+^4}{\ell^2}-q^2\big)\big(1-\frac{r_+^2}{\ell^2}\big)}}{
   1+\frac{6r_+^2}{\ell^2}-\frac{3r_+^4}{\ell^4}-\frac{q^2}{\ell^2}}
\ee
This is the same expression as the one for the central charge
characterizing the chiral CFT dual to the near-horizon geometry of 
the {\em extremal} Kerr-Newman-AdS-dS black hole evaluated in \cite{HMNS0811}.

As in the case of the near-NHEK geometry considered in \cite{Ras1004}, we find that
the central charge $c_R$ associated with the conserved charges $Q_\zeta$ vanishes.
By evaluating $\sqrt{-g}k_\xi[\mathcal{L}_\zeta g;g]$ and 
$\sqrt{-g}k_\zeta[\mathcal{L}_\xi g;g]$ explicitly, it is also verified that there is no central charge
mixing the two conformal sectors characterized by the central charges $c_L$ and $c_R$. 
This establishes that the two corresponding Virasoro algebras are mutually commutative
and suggests that the CFT dual to the {\em near-extremal} Kerr-Newman-AdS-dS black hole
is {\em non-chiral}.
Since $c_R=0$, the Cardy formula for the entropy of this dual CFT is simply given by
\be
 S=\frac{\pi^2}{3}c_LT_L=\frac{\pi(r_+^2+a^2)}{\Xi}
\ee
This is readily recognized as the entropy (\ref{TH}) of the Kerr-Newman-AdS-dS black hole
in the \mbox{(near-)extremal} limit.


\subsection*{Acknowledgments}
\vskip.1cm
\noindent
This work is supported by the Australian Research Council. 
The author thanks Omar Foda for discussions.


\end{document}